\documentclass[runningheads]{svmult}
\usepackage{psfig}
\usepackage{graphicx}
\def\lsim{\lower.5ex\hbox{$\; \buildrel < \over \sim \;$}}
\def\gsim{\lower.5ex\hbox{$\; \buildrel > \over \sim \;$}}
\def\ch{\lower-0.55ex\hbox{--}\kern-0.55em{\lower0.15ex\hbox{$h$}}}
\def\lh{\lower-0.55ex\hbox{--}\kern-0.55em{\lower0.15ex\hbox{$\lambda$}}}
\def\lsim{\lower.5ex\hbox{$\; \buildrel < \over \sim \;$}}
\def\gsim{\lower.5ex\hbox{$\; \buildrel > \over \sim \;$}}
\def\AHR{analogue Hawking radiation}

\begin{document}

\title*{Pseudo-Schwarzschild Spherical Accretion as a Classical Black Hole Analogue}

\author{
     Surajit Dasgupta\inst{1}
\and Neven Bili\'c\inst{2}
\and Tapas K. Das\inst{3}
}

\authorrunning{Dasgupta et al.}

\institute{
Department of Astronomy and Astrophysics,
Tata Institute of Fundamental Research, Mumbai-400005, India.
email: surajit@tifr.res.in
\and Rudjer Bo\v{s}kovi\'{c} Institute, 10002 Zagreb, Croatia
email:  bilic@thphys.irb.hr
\and Theoretical Physics Division, Harish Chandra Research 
Institute, Allahabad-211 019, India.
email: tapas@mri.ernet.in
}

\maketitle

\begin{abstract}
We demonstrate that a spherical accretion onto astrophysical 
black holes, under the influence of Newtonian or various
post-Newtonian pseudo-Schwarzschild gravitational potentials,
may constitute a concrete example of classical analogue gravity 
naturally found in the Universe. We analytically calculate the 
corresponding analogue Hawking temperature as a function of 
the minimum number of physical parameters governing the accretion 
flow. We study both the polytropic and the isothermal accretion. 
We show that unlike in a general relativistic spherical accretion, 
analogue white hole solutions can never be obtained in such 
post-Newtonian systems. We also show that an isothermal spherical 
accretion is a remarkably simple example in which the  only one 
information---the temperature of the fluid, is sufficient to  
completely describe an analogue gravity system. For both types 
of accretion, the analogue Hawking temperature may become higher 
than the usual Hawking temperature. However, the analogue Hawking 
temperature for accreting astrophysical black holes is considerably 
lower compared with the temperature of the accreting fluid.
\end{abstract}


\section{Introduction}
\subsection{Analogue gravity and analogue Hawking temperature}
\noindent
Classical black-hole analogues  are  dynamical fluid
analogues of general relativistic black holes. Such analogue gravity effects may be  observed when
acoustic perturbations propagate through a classical
transonic fluid.
At the transonic point of the fluid flow
the so-called {\em acoustic horizon} is formed.
The acoustic horizon
 resembles a black-hole event horizon
in many ways. In particular,
the acoustic horizon emits
a quasi-thermal phonon spectrum similar to Hawking radiation.

In his pioneering work Unruh (1981) showed
  that the scalar field representing  acoustic perturbations
in a transonic barotropic irrotational fluid,
satisfied a differential equation of the same form as
the Klein-Gordon equation for the massless scalar field propagating in  curved
space-time with a metric that
closely resembles the Schwarzschild metric near the horizon.
 The acoustic
propagation through a transonic fluid forms an analogue event horizon
 located at the transonic point.
Acoustic waves with a quasi-thermal
spectrum will be emitted from the acoustic horizon and the temperature of such
acoustic emission may be calculated as (Unruh 1981)
\begin{equation} T_{\rm ah}=\frac{\hbar}{4{\pi}k_{\rm B}}\left[\frac{1}{a_s}\frac{{\partial}{u_{\perp}^2}}{\partial{n}}
\right]_{\rm ah}\, ,
\label{(1)}
\end{equation}
where
$k_{\rm B}$ is  Boltzmann's
constant,
$a_s$  the speed of sound,  $u_{\perp}$  the component of the
flow velocity normal to the acoustic horizon and
$ \partial/\partial n$
represents the normal derivative.
The subscript `ah' denotes that the quantity should be evaluated at the
acoustic horizon.
The temperature $T_{\rm ah}$ defined by (\ref{(1)}) is an acoustic
analogue of the usual
Hawking temperature $T_{\rm H}$:
\begin{equation}
T_{\rm H}=\frac{{\hbar}c^3}{8{\pi}k_{\rm B}GM_{\rm bh}}
\label{(2)}
\end{equation}
and hence
$T_{\rm ah}$
is referred to as
the {\em analogue  Hawking temperature} or simply
{\em analogue temperature}.
Similarly, the
radiation from the acoustic (analogue) black hole is dubbed {\em analogue Hawking
radiation}. Note that the sound speed in equation (\ref{(1)}) in Unruh's original
treatment  was assumed constant in space.

Unruh's work was followed by other important papers (Jacobson 1991,
Unruh 1995, Visser 1998, Jacobson 1999, Bili\'c  1999).
As shown by
Visser (1998),
 the equation of motion for the
 acoustic disturbance in a barotropic, inviscid fluid
 is identical to the
d'Alembertian equation of motion for a minimally coupled massless scalar field
propagating in (3+1) Lorenzian geometry. In particular, the
acoustic metric for a point sink was shown to be
conformally related to the
Painlev\'e-Gullstrand-Lema\^\i{}tre  form of the Schwarzschild metric. The corresponding expression for the analogue temperature was
shown to be (Visser 1998)
\begin{equation}
T_{\rm ah}=\frac{{\hbar}}{4{\pi}k_{\rm B}}
\left[\frac{1}{a_s}\frac{\partial}{\partial{n}}
\left(a_s^2-u_{\perp}^2\right)\right]_{\rm ah} \, .
\label{(3)}
\end{equation}
Hence, to determine the
analogue Hawking temperature of a classical analogue system,
one needs to know the location of the
acoustic horizon and the values of the speed of sound, the fluid velocity
and their space gradients on the acoustic horizon.

In the analogue gravity systems discussed above, the fluid flows in
 flat Minkowski space, whereas the sound wave
propagating through the non-relativistic fluid is coupled to a curved
pseudo-Riemannian metric. Phonons are
null geodesics, which generate a null surface---the acoustic horizon.
Introduction of viscosity may destroy  Lorenz invariance and hence
the acoustic analogue is best studied in a vorticity free
dissipationless fluid.
\subsection{Motivation: Transonic black-hole accretion as an analogue system}
\noindent
Astrophysical black holes are the final stage of gravitational
collapse of massive celestial objects. Astrophysical black holes may be roughly
classified in two categories:  stellar mass black holes having a mass
of a few $M_{\odot}$
and supermassive black holes of mass
of the order of
${10^6}M_{\odot}$ or more,
where $M_{\odot}$ denotes the solar mass.
Both the stellar mass and the supermassive astrophysical black holes
accrete matter from the surroundings. Depending on the intrinsic
angular momentum of the accreting material, either
the spherically symmetric
(with zero angular momentum)  or
the axisymmetric (with non-zero angular momentum)
 geometry is invoked to study  accreting black-hole systems
(for details see, e.g. Frank, King \& Raine 1992).

Denoting by $u(r)$ and $a_s(r)$  the dynamical velocity and the local sound speed
 of the accreting fluid moving along a space curve parameterized by $r$,
 the local Mach number of the
 fluid is defined as $M(r)=u(r)/a_s(r)$.
The flow  is  locally
 {\em subsonic} or {\em supersonic} if $M < 1$  or $ M>1$, respectively.
 The flow is said to be {\em transonic} if there exists
 at least one point $r_h$ such that
 $M(r_h)=1$. At this point a subsonic to a supersonic
 transition takes place either continuously or discontinuously.
The point where this crossing
takes place continuously is called {\em sonic point},
and if this crossing is not continuous,
 the transition point is called {\em shock}
or {\em discontinuity}.
As a consequence of the inner boundary conditions imposed by the
event horizon, the accretion onto black holes is generally
transonic.
The investigation of accretion processes onto celestial objects
was initiated by
Hoyle \& Littleton (1939) who computed the rate at which
pressureless matter would be captured by a moving star.
 Subsequently,
a theory of
stationary, spherically symmetric transonic hydrodynamic accretion
onto a gravitating astrophysical object at rest was
formulated  by Bondi (1952) using a purely Newtonian approach that
includes pressure effects of the accreting material.

Since the publication of the seminal paper by Bondi in 1952,
the transonic
behaviour of the accreting fluid onto compact astrophysical objects has
been extensively studied in the astrophysics community.
Similarly,
 the
pioneering work by Unruh in 1981
initiated a substantial number of works
in the theory of {\AHR}  with diverse fields of application
(for a review, see Novello, Visser \& Volovik 2002).
Despite the fact that the
accreting black hole can be considered a natural
example of an analogue gravity system,
 an attempt to bridge
these two fields of research has been made
only recently
(Das 2004).
  Since both the theory of transonic
astrophysical accretion and the theory of
{\AHR}  are related by the  same physical phenomena,
it is important to study
the {\AHR} in a transonic accretion onto astrophysical black
holes and to compute $T_{{\rm ah}}$.
 An accreting black-hole system
  of classical analogue gravity is a unique example of classical analogue gravity which
 exhibits
both the black-hole event horizon and  the analog acoustic horizon.
Hence, an accreting astrophysical black hole may  be
considered  an ideal candidate to theoretically
study these two different types of horizons
and to compare their properties.

\subsection{Pseudo-Schwarzschild black-hole accretion}
\noindent
Since the
relativistic effects play an important role in the
regions  close to the accreting black hole,
a purely Newtonian gravitational
potential in the form ${\Phi}_{\rm N}=-GM/r$
is certainly not  a realistic choice to describe
transonic black-hole accretion in general.
However, a rigorous investigation of transonic black-hole accretion (even
with spherical symmetry) may be  quite
complicated in  a general relativistic
treatment.
In order to compromise between a relatively easy
handling of the
Newtonian description of gravity and realistic but
complicated general relativistic calculations, a series of
`modified' Newtonian potentials have been introduced.
The following potentials have been proposed:
\begin{eqnarray}
\Phi_{1}=-\frac{1}{2(r-1)},\;\;\;\;\;\;
 \Phi_{2}=-\frac{1}{2r}\left[1-\frac{3}{2r}+
12{\left(\frac{1}{2r}\right)}^2\right] ,  && \nonumber \\
\Phi_{3}=-1+\left(1-\frac{1}{r}\right)^{1/2},\;\;\;\;\;\;
\Phi_{4}=\frac{1}{2}\ln {\left(1-\frac{1}{r}\right)} \, ,
\;\;\; \;\;\;\;\;\; &&
\label{(4)}
\end{eqnarray}
where the units
$G=c=M_{\rm bh}=1$ have been used.
Unless otherwise stated, hereafter we use these units.
The radial distance $r$ in (\ref{(4)}) is measured in units of
the  Schwarzschild gravitational radius
$r_g=2G M_{\rm bh}/c^2$, where  $M_{\rm bh}$  is the mass of the black
hole, $G$  Newton's gravitational constant and $c$ is the
speed of light.
The potential $\Phi_1$ is proposed by Paczy\'nski and Wiita (1980),
$\Phi_2$ by Nowak and Wagoner (1991), $\Phi_3$ and
$\Phi_4$ by Artemova, Bj\"{o}rnsson \&
Novikov (1996).

The use of these potentials
retains most of the features of space-time around a compact object
up to a reasonably small distance from the black-hole
event horizon, and
helps in reproducing
some crucial properties of the relativistic
accretion solutions with high accuracy.
Hence, these potentials might be called {\em post-Newtonian
pseudo-Schwarzschild potentials} (for details, see Artemova, Bj\"{o}rnsson \&
Novikov 1996 and Das 2002).
These potentials were originally introduced to study
the axisymmetric accretion (accretion discs). However, it was shown
(Das \& Sarkar 2001) that some of these potentials could efficiently describe
a spherically symmetric accretion as far as comparison with the full
general relativistic accretion is concerned.

In this paper we  study the analogue gravity phenomena in the spherical
accretion onto astrophysical black holes under the influence of various post-Newtonian
pseudo-Schwarzschild potentials described above.
We  use the expressions `post-Newtonian' and `pseudo-Schwarzschild'
synonymously.
Our main goal is to provide a self-consistent calculation of
the analogue horizon temperature $T_{\rm ah}$
 in terms of the minimum number of
physical accretion parameters, and to study the dependence of $T_{\rm ah}$
on various flow properties.
In most practical situations, a complete general relativistic treatment
turns out to be almost non-tractable.
In those few simple cases where  the  general relativistic treatment is easy,
 one can repeat the calculations using
 various pseudo-potentials, and can compare the results with those obtained in
the  general relativistic calculations.
This comparison may help to handle more complicated accretion systems
without resorting to the full general relativistic treatment.

In Section 2 we calculate the location of the
acoustic horizon and evaluate the
 relevant accretion variables on the horizon as  functions
of the fundamental parameters governing the flow.
Hereafter, we use the expressions
`acoustic horizon' and `analogue horizon' synonymously.
We study both the polytropic
(adiabatic) and the isothermal accretion. In Section  3 we  derive an analytic expression for
$T_{\rm ah}$ as a function of various accretion parameters. In Section 4 we present
our results numerically and in Section 5 we give a discussion and conclusions.
\section{Calculation of various sonic quantities}
\noindent
The non-relativistic equation of motion  for spherically accreting matter
in a gravitational potential denoted by $\Phi$
 may be written as
\begin{equation}
\frac{{\partial{u}}}{{\partial{t}}}+u\frac{{\partial{u}}}{{\partial{r}}}+\frac{1}{\rho}
\frac{{\partial}p}{{\partial}r}+\frac{{\partial}\Phi}{\partial{r}}=0    ,
\label{(5)}
\end{equation}
where $u$, $p$ and $\rho$ are the velocity, pressure and density of the fluid, respectively.
The first term in equation (\ref{(5)}) is the Eulerian time derivative of the
dynamical velocity, the second term
is the `advective' term, the third term
is the
momentum deposition due to the pressure gradient and the
last term  is the gravitational force.
Another equation necessary to describe
the motion of the fluid is
the continuity
equation
\begin{equation}
\frac{{\partial}{\rho}}{{\partial}t}+\frac{1}{r^2}\frac{{\partial}}{{\partial}r}\left({\rho}ur^2\right)=0 .
\label{(6)}
\end{equation}
To integrate the above set of equations, one also needs the
equation of state that specifies the intrinsic properties of the fluid.
 As mentioned earlier, we will study  accretion described by either a polytropic
or an isothermal equation of state.
\subsection{Polytropic accretion}
We employ
a polytropic equation of state of the form
\begin{equation}
p=K{\rho}^\gamma ,
\label{(7)}
\end{equation}
 where
 the polytropic index  is equal to the ratio of two
specific heats, $\gamma=c_p/c_V$.
The  sound speed $a_s$ is defined by
\begin{equation}
a_s^2\equiv\left. \frac{\partial p}{\partial\rho}\right|_s=\gamma\frac{p}{\rho},
\label{(8)}
\end{equation}
where the subscript $s$ denotes that the derivative is taken while keeping
 the  entropy per particle
$s=S/n$ fixed, i.e. for an isentropic process.
Assuming stationarity of the flow, we find the following conservation equations:

\noindent
1) Conservation of energy  implies  constancy
of the specific energy ${\cal E}$
\begin{equation}
{\cal E}=\frac{u^2}{2}+\frac{a_s^2}{{\gamma}-1}+\Phi .
\label{(9)}
\end{equation}

\noindent
2) Conservation of the baryon number implies constancy of the accretion rate ${\dot M}$
\begin{equation}
{\dot M}=4{\pi}{\rho}ur^2 .
\label{(10)}
\end{equation}
Equation (\ref{(9)}) is obtained  from (\ref{(5)}) with (\ref{(7)}) and (\ref{(8)})
and equation (\ref{(10)}) follows directly from (\ref{(6)}).

Substituting $\rho$ in terms of $a_s$ and
differentiating (\ref{(10)}) with respect to $r$,
 we obtain
\begin{equation}
a_s'=\frac{a_s (1-\gamma)}{2}
\left(\frac{u'}{u}+\frac{2}{r}\right) ,
\label{(11)}
\end{equation}
where $'$ denotes the derivative with respect to $r$.
Next we  differentiate equation (\ref{(9)}) and eliminating $a_s'$ with the help
of (\ref{(11)}) we obtain
\begin{equation}
u'=\frac{2a_s^2/r-
\Phi'}{u-a_s^2/u} \, .
\label{(12)}
\end{equation}
If the denominator  of equation (\ref{(12)})  vanishes at a particular radial distance
$r_h$, the numerator must also vanish at $r_h$  to maintain the
continuity of the flow.
One thus finds the sonic-point condition
as
\begin{equation}
{u}_h=a_{sh}=\sqrt{\frac{r_h\Phi'_h}{2}} \, ,
\label{(13)}
\end{equation}
where $r_h$ is the sonic point and the sphere of radius $r_h$ is the
acoustic horizon.
Hereafter, the subscript $h$ indicates that a particular
quantity is evaluated at $r_h$.
The location of the acoustic horizon
is obtained
by solving the algebraic equation
\begin{equation}
{\cal E}-\frac{1}{4}\left(\frac{\gamma+1}{\gamma-1}\right)r_h
{\Phi'_h}
-{\Phi_h}=0 .
\label{(14)}
\end{equation}
The derivative  $u'_h$
at the
corresponding sonic point is obtained by
solving the quadratic equation
\begin{eqnarray}
\left(1+\gamma\right)\left(u'_h\right)^2+
2\left(\gamma-1\right)\sqrt{\frac{2\Phi'_h}{r_h}}\,
u'_h & &
\nonumber \\
+\left(2{\gamma}-1\right)
\frac{\Phi'_h}
{r_h}
+{\Phi''_h} &= & 0,
\label{(15)}
\end{eqnarray}
which follows from (\ref{(12)}) in  the limit
$r{\rightarrow}r_h$
 evaluated with the help of l'Hospital's rule.
 
Finally, the gradient of the sound speed
at the acoustic horizon is obtained
by substituting  $u'_h$ obtained from (\ref{(15)})
into equation (\ref{(11)}) at the acoustic horizon
\begin{equation}
\left. a_s'\right|_h=\left(\frac{1-\gamma}{2}\right)
\left(u'_h+\sqrt{\frac{2\Phi'_h}{r_h}}\right) .
\label{(16)}
\end{equation}
Evidently, for both the Newtonian flow
with $\Phi_{\rm N}=-1/r$ and the post-Newtonian flow, the
location of the sonic point is identical to the location of the
acoustic horizon due to the assumption of stationarity and spherical symmetry.

\subsection{Isothermal Accretion}
\noindent
We employ the isothermal equation of state of the form
\begin{equation}
p=\frac{RT}{\mu}\rho=c_s^2{\rho}\, ,
\label{(17)}
\end{equation}
where $T$ is the
temperature,
$R$ and $\mu$ are the universal gas constant and the mean molecular weight, respectively.
 The quantity  $c_s$ is the isothermal sound speed defined by
\begin{equation}
c_s^2=\left.\frac{\partial p}{\partial \rho}\right|_T
={\Theta}T \, ,
\label{(18)}
\end{equation}
where the derivative is taken at fixed  temperature and the constant
$\Theta=\kappa_b/(\mu m_H)$ with $m_H \simeq m_p$ being the
mass of the hydrogen atom.
In our model we assume that the accreting matter is predominantly hydrogen,
hence $\mu \simeq 1$.
Now, the specific energy equation takes the form
\begin{equation}
{\cal E}=\frac{u^2}{2}+{\Theta}T\ln \rho+\Phi \, ,
\label{(19)}
\end{equation}
whereas the accretion rate is given by (\ref{(10)}) as before.

The radial change rate of the dynamical velocity
is again given
 by (\ref{(12)}). From equation (\ref{(12)})  and with (\ref{(18)})
we find
the sonic point condition as
\begin{equation}
u_h=\sqrt{\frac{r_h\Phi'_h}{2}}
=c_s=\sqrt{{\Theta}T} \, .
\label{(20)}
\end{equation}
 The derivative of $u$ is obtained
from (\ref{(12)})
by making use of l'Hospital's rule as before. We find
\begin{equation}
u'_h=-
\sqrt{-\frac{1}{2}\left(\frac{1}{r_h}\Phi'_h+\Phi''_h\right)}\, ,
\label{(21)}
\end{equation}
where the minus sign in front of the square root
indicates accretion (the plus would
correspond to a wind solution).
Note that the quantities in equations (\ref{(20)}) and (\ref{(21)}) are functions of
the fluid
temperature $T$  only. Hence the isothermal spherical accretion can be
essentially described as a one-parameter solution of the
hydrodynamical equations, parameterized by $T$.
\section{Calculation of Analogue Temperature}
\noindent
To calculate $T_{\rm ah}$ for various $\Phi$, we first write down the general
expression for $T_{\rm ah}$ in the relativistic form and then we reduce the
expression by taking the non-relativistic post-Newtonian limit.

 The relativistic
acoustic metric tensor $G_{\mu{\nu}}$ is defined as
(Moncrief 1980, Bili\'c 1999)
\begin{equation}
G_{\mu{\nu}}=\frac{n}{{h}{a_s}}\left[g_{\mu{\nu}}-\left(1-a_s^2\right)
u_{\mu}u_{\nu}\right]  \, ,
\label{(22)}
\end{equation}
where $n$ is the
particle number density and $h=(p+\rho)/n$ is the specific enthalpy.
The corresponding surface gravity may be calculated as (Bili\'c 1999)
\begin{equation}
\kappa=\left.\frac{\sqrt{{\xi}^{\nu}{\xi}_{\nu}}}{1-a_s^2}
\frac{\partial}{\partial{n}}
\left(u-a_s\right)\right|_{\rm ah} \, ,
\label{(23)}
\end{equation}
where $\xi^{\mu}$ is the stationary Killing field and $\partial/{\partial{n}}$ is the
derivative normal to the acoustic horizon.
This expression together with
the Newtonian limit
\begin{equation}
|\xi^2|=g_{00}\rightarrow{\left(1+\frac{\Phi}{2c^2}\right)}
\label{(24)}
\end{equation}
gives a general expression for the
temperature of the analogue Hawking radiation in a spherically
accreting fluid
 in the Newtonian as well as in any
pseudo-Schwarzschild gravitational potential
\begin{equation}
T_{\rm ah}=\frac{\hbar}{2{\pi}\kappa_b}
\sqrt{\frac{2c^2+\Phi_h}{2c^2}}
\left[\frac{1}{1-a_s^2}\left|\frac{d}{dr}
\left(a_s-u\right)\right|\right]_{\rm ah} \, .
\label{(25)}
\end{equation}
The quantities required to calculate the analogue temperature (\ref{(25)}) are
obtained using the formalism presented in Section 2.1.
The Newtonian and post-Newtonian polytropic
spherical accretion onto astrophysical black holes is an example
of an analogue-gravity system in which the  value of $T_{\rm ah}$ can be calculated
using only two measurable physical parameters, namely, ${\cal E}$ and $\gamma$.

For a particular value of $\{ {\cal E}, \gamma\}$, we define a dimensionless quantity $\tau$ as the ratio of
$T_{\rm ah}$ to $T_{\rm H}$
which turns out
 to be independent of the black-hole mass.
In this way we can
compare the properties of the acoustic versus event horizon of
a spherically
accreting black hole of any mass.
Using equations (\ref{(11)})-(\ref{(16)}) we  find
\begin{eqnarray}
\tau\equiv \frac{T_{\rm ah}}{T_{\rm H}} =
4\sqrt{\frac{2+\Phi_h}{2}}\left(\frac{2}{2-r_h\Phi_h}\right)
\left(\frac{\gamma+1}{2}\right)
\nonumber  \\
\sqrt{\frac{\Phi'_h}{r_h}{\bf f}(\gamma)-
\left(1+\gamma\right)\Phi''_h}
\label{(26)}
\end{eqnarray}
where ${\bf f}(\gamma)=\left(0.00075\gamma^2-5.0015{\gamma}+3.00075\right)$.
The quantities
$\Phi_h$, $\Phi'_h$, and $\Phi''_h$
are obtained from
 equation (\ref{(4)}) for various potentials,
and  $r_h$ is calculated from (\ref{(14)}) for an
astrophysically
relevant choice of $\{ {\cal E}, \gamma\}$.

The quantity ${\cal E}$ is scaled in terms of
the rest mass energy. The values ${\cal E}<0$ correspond to negative energy accretion states where a
radiative extraction of the
rest mass energy from the fluid is required.
To make  such an extraction
 possible, the accreting fluid has to
possess viscosity or other dissipative mechanisms, which may violate
 Lorenz invariance.
The values ${\cal E}>1$ are mathematically allowed. However,  large positive
${\cal E}$ represent the flows  that start from infinity
with a very high thermal energy, and the ${\cal E} > 1$ accretion represents enormously
hot flow configurations at very large distances from the black hole.
These configurations
are not  conceivable in realistic astrophysical situations.
Hence we set  $0{\lsim}{\cal E}{\lsim}1$.

Concerning the polytropic index,
the value $\gamma=1$ corresponds to the isothermal accretion
in which the accreting fluid remains optically thin.
 This value is the physical lower limit  since
 $\gamma<1$ is not realistic in accretion
astrophysics. On the other hand,
$\gamma>2$ is possible only for a superdense matter
with a substantially large magnetic
field
and a direction-dependent anisotropic pressure.
This requires the accreting material to be governed by
magneto-hydrodynamic
equations, dealing with which
is beyond the scope of this paper.  We  thus
set $1{\lsim}\gamma{\lsim}2$ for a  polytropic flow, and
separately consider the case $\gamma=1$, corresponding  to an
isothermal accretion.
The  preferred
values of $\gamma$ for a realistic
 polytropic black hole accretion range from $4/3$
to $5/3$ (Frank et. al. 1992).

Note that if $(a_s' - u')_h$ is negative, one obtains
an acoustic  {\it white-hole} solution.
Hence the condition for the existence of the acoustic white hole is
\begin{equation}
\left(\frac{\gamma+1}{2}\right)
\sqrt{\frac{\Phi'_h}{r_h}{\bf f}(\gamma)-
\left(1+\gamma\right)\Phi''_h}\: < 0 .
\label{(27)}
\end{equation}
Since $\gamma$ and $r_h$ can never be negative, and since
$\Phi''_h$ and $\Phi''_h$ are
always real for the preferred domain of $\{ {\cal E}, \gamma\}$,
 acoustic white-hole
solutions are excluded in the astrophysical accretion governed by the Newtonian or
post-Newtonian potentials.

For an isothermal flow, the quantity
$c_s'$ is zero and
using (\ref{(21)}) we find
\begin{equation}
\tau=4\sqrt{2}\left(\frac{1}{2-r_h\Phi'_h}\right)
\sqrt{-\left(1+\frac{\Phi_h}{2}\right)
\left(\Phi''_h+\frac{\Phi'_h}{r_h}\right)} \, ,
\label{(28)}
\end{equation}
where $r_h$ should be evaluated using (\ref{(20)}).
Clearly,  the fluid temperature $T$ completely determines 
the analogue Hawking temperature. Hence, a spherical 
isothermally accreting astrophysical black hole provides 
a simple system where  analogue gravity can be theoretically 
studied using only one free parameter.

\begin{figure}
\vbox{
\vskip 0.5cm
\centerline{
\psfig{file=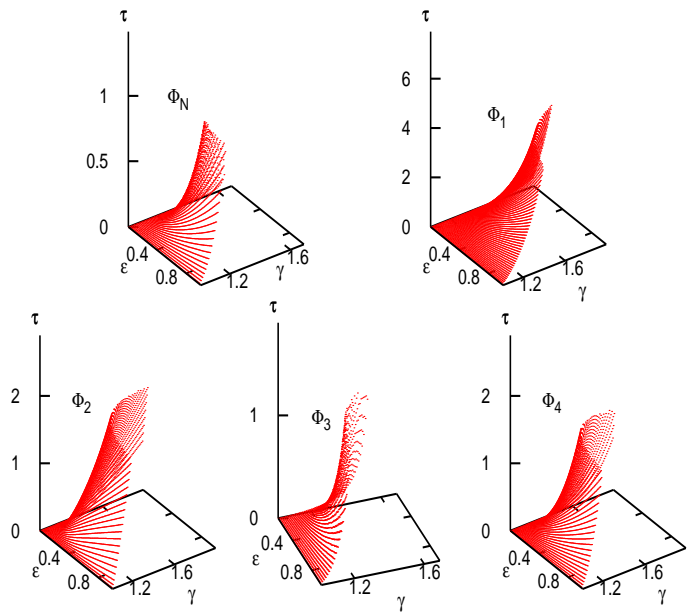,height=8cm,width=9cm,angle=0.0}}}
\caption{
The ratio of the analogue to
the Hawking temperature $\tau=T_{\rm ah}/T_{\rm H}$ as a function of the
specific flow energy ${\cal E}$ and the
polytropic index $\gamma$
for the  Newtonian (marked with $\Phi_{\rm N}$)
and four pseudo-Schwarzschild potentials (marked with $\Phi_1$, $\Phi_2$, $\Phi_3$
and $\Phi_4$).
}
\end{figure}
\section{Results}
\subsection{Polytropic accretion}
\noindent
Given a potential $\Phi$, we calculate $r_h$ as a function of
  $\{{\cal E},\gamma\}$
 using equation (\ref{(15)}).
The physically motivated choice of the domain $\{ {\cal E}, \gamma\}$  is discussed in Section 3.
 Note that the physically acceptable values of $r_h$
must be larger than 1
since
 the black-hole event horizon is located at $r=1$.
However, we show the results only for $r_h{\ge}2$ since the
pseudo-potentials (\ref{(4)}) may not
 accurately mimic the Schwarzschild
space-time too  close to the
black-hole event horizon.
We find that $r_h$ anti-correlates with $\{ {\cal E}, \gamma\}$,  which  is
obvious from equation (\ref{(14)}).
Physically, this effect is due to a subtle competition between
the magnitudes and radial gradients of the velocity field  and
the speed of sound.
On the one hand, increase of ${\cal E}$ at fixed
$\gamma$
 means increase
of the asymptotic energy configuration of the flow which is basically
thermal in nature. As a result,
the larger
 the energy ${\cal E}$, the closer to the accretor must the flow
approach in order to gain sufficiently large dynamical velocity  and its gradient.
 On the other hand, keeping ${\cal E}$ fixed,
 increase of $\gamma$ means  increase of the heat capacity of the
accreting fluid and in turn
 its speed of sound.
 As a result,
the flow becomes supersonic at a smaller distance to the black hole.

Once we have calculated $r_h$,
  we calculate the ratio  $\tau= T_{\rm ah}/T_{\rm H}$
as a function of two flow parameters ${\cal E}$ and $\gamma$
using (\ref{(26)})
and plot the results in figure 1 for various potentials.
\begin{figure}
\vbox{
\vskip 0cm
\centerline{
\psfig{file=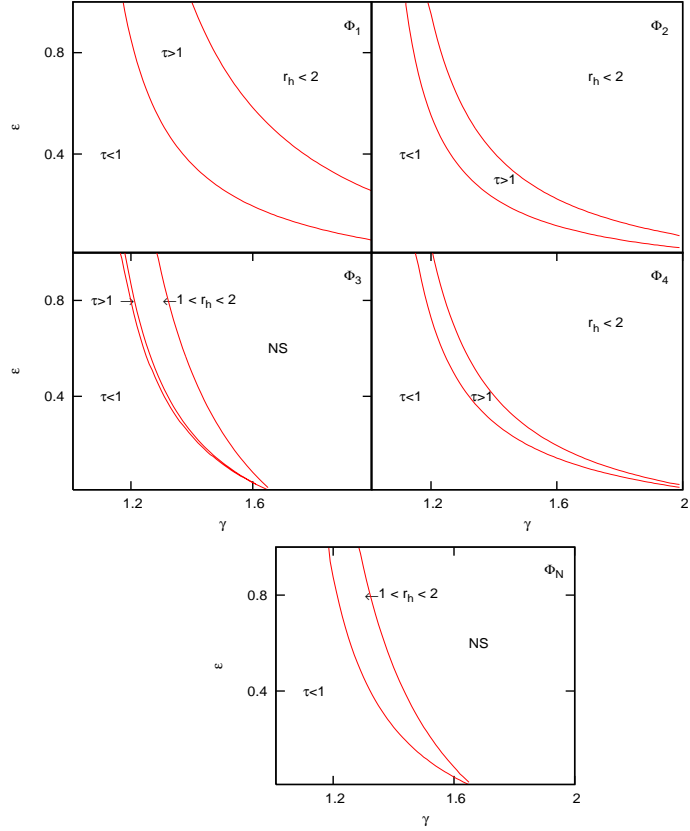,height=11cm,width=9cm,angle=0.0}}}
\caption{
Classification of the $\{ {\cal E}, \gamma\}$ parameter space for a polytropic spherical accretion
flow in the Newtonian (marked with $\Phi_{\rm N}$) and in four pseudo-Schwarzschild potentials
(marked with $\Phi_1$, $\Phi_2$, $\Phi_3$ and $\Phi_4$).
The parameter space is divided in four regions in which
the analogue temperature $T_{\rm ah}$ is always lower than the
Hawking temperature $T_{\rm H}$ (marked with $\tau < 1$),
$T_{\rm ah}$ is higher than $T_{\rm H}$ (marked with $\tau > 1$),
the use of the pseudo-potentials is unreliable (marked with  $1<r_h<2$),
the accretion flow is always subsonic (marked with {\bf NS}).
}
\end{figure}
Note that for all potentials except $\Phi_{\rm N}$,
the ratio
$\tau$ may be larger than unity for some $\{ {\cal E}, \gamma\}$ and increases
both with increasing ${\cal E}$ and with increasing $\gamma$.
In the region
 $1<r_h < 2$, the ratio $\tau$ may become very large, and
a small zone may appear
where $\tau$ is again less than one.
In the Newtonian case, $\tau$  becomes larger than unity
only in the region $1<r_h<2$.
However, as discussed
earlier, we do not rely on the results obtained in this region.

In figure 2 we classify the $\{ {\cal E}, \gamma\}$ parameter space to show some distinct important
regions. The region marked with  $\tau < 1$ is a range of values of
$\{ {\cal E}, \gamma\}$ for which the analogue temperature is always lower than the
Hawking temperature.
Similarly,
$\tau > 1$ denotes a range of values of
$\{ {\cal E}, \gamma\}$ for which $T_{\rm ah}$ is higher than $T_{\rm H}$.
In these regions we have
$r_h{\ge}2$. The region marked with $1<r_h<2$
represents the values of $\{ {\cal E}, \gamma\}$ for which
 a physical solution may be obtained
for
$1<r_h<2$, and $\tau$ may be larger than (mostly) and
less than (rarely) unity.
The regions marked with  {\bf NS} represent the
$\{ {\cal E}, \gamma\}$ subset for which no physical
solution $r_h > 1$ exists. Hence, in this region of the parameter space
there is no transonic flow and the acoustic horizon does not form.
Note that such a situation may occur only with
$\Phi_3$ and with purely Newtonian potentials. With all other potentials,
 physical
values ($r_h >1$) exist in the
 entire domain of $\{ {\cal E}, \gamma\}$ .

The domination of the analogue temperature over the
Hawking temperature is  most prominent  with
the Paczy\'nski and Wiita (1980)
potential, and least prominent  with the pure Newtonian potential.
If we denote by $\tau^{\Phi}$  the analogue to the Hawking temperature ratio for
a particular $\Phi$ at a particular value of $\{ {\cal E}, \gamma\}$ for which a
physical solution $r_h$ exists for all five potentials, then
\begin{equation}
\tau^{\Phi_1}~>~\tau^{\Phi_2}~>~\tau^{\Phi_3}~>~\tau^{\Phi_4}~>~\tau^{\Phi_{\rm N}} .
\label{(29)}
\end{equation}

\begin{figure}
\vbox{
\vskip -4.5cm
\centerline{
\psfig{file=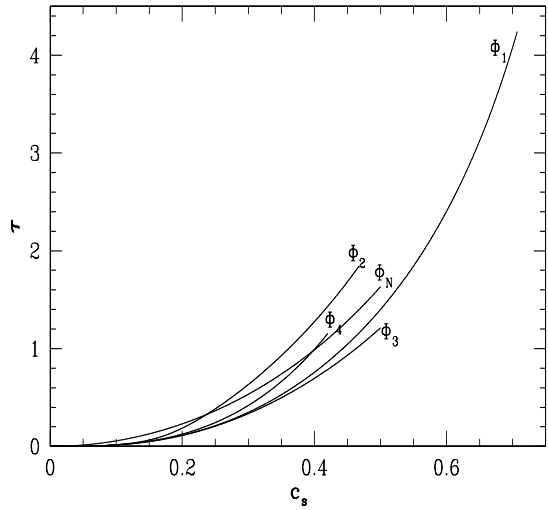,height=12cm,width=9cm,angle=0.0}}}
\caption{
The ratio of the analogue to
the Hawking temperature $\tau=T_{\rm ah}/T_{\rm H}$ as a function of the
acoustic velocity $c_s$ (scaled in the unit of the velocity of light)
at $r_h$, for spherical isothermal accretion
in the Newtonian (marked with $\Phi_{\rm N}$) and four pseudo-Schwarzschild potentials
(marked with $\Phi_1$, $\Phi_2$, $\Phi_3$ and $\Phi_4$).
 Results are shown for
$r_h{\ge}2$ only.
}
\end{figure}
\subsection{Isothermal accretion}
\noindent
As demonstrated in Section 2.2,  the
flow temperature $T$ fully determines the location of the acoustic horizon
$r_h$ and other relevant quantities evaluated at $r_h$.
We take a representative fluid temperature (in degrees of K) in the range $10^8{\le}T{\le}4{\times}10^{12}$, where the
upper limit ensures that $r_h$ should never be less than 2.
We find that
the horizon radius $r_h$
calculated using equation (\ref{(20)})
 increases with $T$. This is obvious because the higher
the temperature, the larger the speed of sound,  and the flow
needs to get closer to the accretor to gain the dynamical velocity
$u$ and become supersonic. For a given temperature $T$, we find
\begin{equation}
r_h^{\Phi_{\rm N}}~>~r_h^{\Phi_2}~>~r_h^{\Phi_4}~
>~r_h^{\Phi_3}~>~r_h^{\Phi_1} \, .
\label{(30)}
\end{equation}
In figure 3 we plot the ratio $\tau$ as a function of the  
isothermal acoustic velocity $c_s$ (scaled in the unit of the velocity of light)
at the acoustic horizon $r_h$,
for five potentials as in figure 1.
The value of $c_s$ is obtained from eq. (20).
The lines in the plot extend as far as a physical value $r_h \ge 2$ exists.
We find that
$\tau$ always increases with $T$ (hence with $c_s$)
and at a particular $T$ (as well as at that corresponding 
$c_s$) we also find
\begin{equation}
\tau^{\Phi_2}~>~\tau^{\Phi_{\rm N}}~>~\tau^{\Phi_4}~>~\tau^{\Phi_1}~>~\tau^{\Phi_3}\, .
\label{(31)}
\end{equation}

\section{Discussion}
\noindent
In this work we have demonstrated that a spherical accretion onto astrophysical
black holes, under the influence of Newtonian and various post-Newtonian
pseudo-Schwarzschild potentials, can be considered as an analogue
gravity system naturally
found in the Universe.
Spherically accreting black holes are unique analogue gravity systems
in which two kinds of horizons,  black-hole event and acoustic analogue
horizon, may be  simultaneously present.
We have calculated
the analogue horizon temperature $T_{\rm ah}$ for spherically accreting systems
as a function of the minimum number of physical parameters governing the
accretion flow.
We have modelled the flow in a self-consistent way
and studied the dependence of $T_{\rm ah}$ on various flow properties.
In our model the calculations have been performed with a
{\em position-dependent}  speed of sound
using  the relativistic formalism  in
curved space (Bili\'c 1999).

Note, however, that the analogy has been applied to
describe the classical perturbation of the fluid in terms of a
field satisfying the wave equation in an effective geometry.
It is not our aim to provide a  formulation by which
the phonon field generated in this system could be quantized.
To
accomplish this task, one would need to show that the effective action for the
acoustic perturbation is equivalent to a field theoretical action
in curved space, and the corresponding commutation and dispersion
relations  should
directly follow (see, e.g. Unruh \& Sch\"utzhold 2003).
Such considerations would be beyond the scope of
this paper.
Also note that for all  types of
accretions discussed here, the analogue temperature $T_{\rm ah}$
is many orders of magnitude  lower compared with the fluid temperature
of accreting matter.
Hence, the black-hole accretion system may not be a good
candidate for detection of the analogue Hawking
radiation.

We have studied both the polytropic and the isothermal flow.
For a polytropic flow, two free parameters ${\cal E}$ and ${\gamma}$ are required
to calculate $T_{\rm ah}$. The polytropic index  $\gamma$ is assumed to be constant throughout
the fluid. A realistic model of the flow with no assumptions
would perhaps require a variable polytropic index having a functional dependence on the radial
distance, i.e. $\gamma=\gamma(r)$ instead of a constant
$\gamma$. However, we  have performed the
calculations for a sufficiently large range of $\gamma$ and we believe
 that all possible astrophysically relevant
polytropic indices are covered.

Das \& Sarkar (2001) have shown that among all pseudo-potentials, $\Phi_1$
and $\Phi_4$
are the best because they provide an accurate approximation
to the full general relativistic solution up to a reasonably
close distance to the event horizon.
 While $\Phi_1$
is the best approximation for an ultra-relativistic flow with
$\gamma=4/3$, the potential
$\Phi_4$ turns out to be the best approximation when the flow
becomes non-relativistic, i.e.
when $\gamma\simeq   5/3$.
Hence, the temperature $T_{\rm ah}$ calculated with
$\Phi_1$ and $\Phi_4$ is comparable with the value of
$T_{\rm ah}$ calculated for a general relativistic flow.
In the isothermal case, there
is presently no such comparative study and hence
no preference among various $\Phi$ may be set for
an isothermal accretion.

From our calculations it follows that the main difference between
 the post-Newtonian polytropic flow
(irrespective of the potential used)
and the full relativistic flow is that
an analogue white hole can never form in the Newtonian or
post-Newtonian spherical accretion,
whereas in a general relativistic accretion,
analogue white-hole solutions exist for
an extended parametric region of $\{ {\cal E}, \gamma\}$ (Das 2004).
In an isothermal accretion with
a constant speed of sound, the question of
 the analogue white-hole formation does not arise at all.

We have found that the
isothermal spherical accretion onto an astrophysical black hole
represents a simple analogue gravity model in  which  one
physical parameter, the  fluid temperature $T$,  completely
specifies the  system and provides sufficient information for the
 calculation of
the analogue temperature.
We would, however, like to raise one point of caution.
 In a realistic astrophysical
situation with perturbations of various types, isothermality of the flow is
 difficult to maintain if
the acoustic horizon is formed at a very large distance from the
event horizon. Owing to the fact that, close to a spherically
accreting Schwarzschild black hole, the electron number density $n_e$ falls off as
$r^{-3/2}$ while the photon number density $n_\gamma$ falls off as $r^{-2}$
(Frank et al. 1992), the ratio of $n_e$ to $n_\gamma$ is proportional to $\sqrt{r}$.
Hence, the
 number of electrons per photon decreases   with decreasing $r$,
so that close to the
accretor, the momentum transfer by photons on the accreting fluid may
 efficiently keep the fluid temperature
roughly constant. Hence, isothermality may  be a
justified assumption at small values of $r_h$.
However, the momentum transfer
 efficiency
falls off with increasing $r$ and
the isothermality assumption
may break down
far away from the black hole.

We have found that in both the adiabatic and the isothermal flow, the analogue temperature
may take over the  Hawking temperature, i.e. the ratio $\tau$
may be larger than unity. This effect has also been
established in a fully relativistic flow (Das 2004).
 One should note that since $\tau$ decreases with increasing $r_h$,
 high values of $\tau$ are obtained if the acoustic horizon is formed
close to the event horizon, i.e. for $1<r_h<2$.
However, our post-Newtonian
model may not be  reliable in this region since
the difference between a post-Newtonian metric  and the general relativistic
one is likely to be
  quite substantial in the vicinity of the  event horizon.
The potentials discussed here
could only be used to obtain a more
accurate description compared with the purely
Newtonian approach.
Hence very large values of $\tau$  are consistent only in a complete
general relativistic treatment of accretion.

One interesting important problem is the study of the analogue gravity
and the calculation of the analogue temperature for a non-spherical axisymmetric accretion (accretion discs).
In this case, the situation is more complicated
because, first,
a realistic disc accretion requires a more involved
parametric space and, second,  more
than one sonic points exist
for specific
 boundary conditions
(Das 2002, Das, Pendharkar \& Mitra 2003,
Das 2004a, Barai, Das \& Wiita 2004).
A detailed study of these issues is beyond the scope of this article and 
is presented elsewhere (Abraham, Bili\'c \& Das).

\section*{Acknowledgments}
The research of SD is partly supported by
the Kanwal Rekhi scholarship of TIFR endowment fund.
TKD acknowledges the hospitality
provided by the Department of Astronomy \& Astrophysics, TIFR.
The work of NB is supported
 in part by the Ministry of Science and Technology of the
 Republic of Croatia under Contract No. 0098002.


\begin{thebibliography}{}
\bibitem{1}
Abraham, H., Bili\'c, N., \& Das, T. K., 2005,
`{\it Acoustic Horizons in Axially Symmetric Relativistic Acretion}',
gr-qc/0509057.
\bibitem{2} Artemova, I. V., Bj\"{o}rnsson,  G., \& Novikov, I. D.  1996, ApJ, 461, 565
\bibitem{3} Barai, P., Das, T. K., \& Wiita, P. J. 2004, Astrophysical Journal Letters,
613, L49.
\bibitem{4} Barcelo, C., Liberati, S., Sonego, S., \& Visser, M. 2004,
New Journal of Physics, Volume 6, Issue 1, pp. 186.
\bibitem{4} Bili\'c, N. 1999, Class. Quant. Grav. 16, 3953
\bibitem{5} Bondi, H. 1952, MNRAS, 112, 195
\bibitem{6} Das, T. K. 2002, ApJ, 577, 880
\bibitem{7} Das, T. K. 2004, Class. Quant. Grav. 21, 5253
\bibitem{8} Das, T. K. 2004a, MNRAS, 375, 384
\bibitem{9} Das, T. K. \& Sarkar, A. 2001, A \& A, 374, 1150
\bibitem{10} Das, T. K., Pendharkar, J. K., \& Mitra, S. 2003, ApJ, 592, 1078
\bibitem{11} Frank, J., King, A. R. \& Raine, D. J. 1992, Accretion Power in Astrophysics
(2nd ed.; Cambridge: Cambridge Univ.Press)
\bibitem{12} Hoyle, F., \& Lyttleton, R. A. 1939, Proc. Camb. Phil. Soc, 35, 59
\bibitem{13} Jacobson, T. A. 1991, Phys. Rev. D. 44, 1731
\bibitem{14} Jacobson, T. A. 1999, Prog. Theor. Phys. Suppl. 136, 1
\bibitem{15} Moncrief, V. 1980, ApJ. 235, 1038
\bibitem{16} Novello, Visser \& Volovik (ed.) 2002, Artificial Black Holes.
World Scientific, Singapore.
\bibitem{17} Nowak, A. M., \& Wagoner, R. V. 1991, ApJ, 378, 656
\bibitem{18} Paczy\'nski, B., \& Wiita, P. J. 1980, A \& A, 88, 23
\bibitem{19} Unruh, W. G. 1981, Phys. Rev. Lett. 46, 1351
\bibitem{20} Unruh, W. G. 1995, Phys. Rev. D. 51, 2827
\bibitem{21} Unruh, W. G. \& Sch$\ddot{\rm u}$tzhold, R. 2003, Phys. Rev. D. 68, 024008
\bibitem{22} Visser, M. 1998, Class. Quant. Grav. 15, 1767
\end{thebibliography}
\end{document}